\documentclass[twocolumn,showpacs,preprintnumbers,amsmath,amssymb,superscriptaddress,prl]{revtex4}
\usepackage[english]{babel}
\usepackage{mathrsfs}
\usepackage{epsfig}
\usepackage{multirow}
\usepackage{graphicx}

\def\p{\partial}
\def\bx{\bgroup \bf x\egroup}

\def\bn{\textbf{n}}
\def\const{{\rm const}}

\def\csch{\mathop{\rm csch}\nolimits}

\def\ve{\varepsilon}
\def\vp{\varphi}

\def\be{\begin{equation}}
\def\ee{\end{equation}}
\def\mymargin#1{\bgroup * \marginpar{*\,#1}}
\let\vec\bf

\RequirePackage{color}
\def\new#1{\bgroup \color{red}{#1}\egroup}
\def\old#1{\bgroup  \it #1\egroup}

\begin{document}


\title{Doubly-periodic instability pattern in a smectic~A liquid crystal}

\author{O. V. Manyuhina}
\email{oksanam@nordita.org}
\affiliation{Nordita, KTH Royal Institute of Technology and Stockholm University, Roslagstullsbacken 23, SE-106 91 Stockholm, Sweden}
\author{G. Tordini}
\affiliation{High Field Magnet Laboratory, Institute for Molecules and Materials, Radboud University Nijmegen
Toernooiveld 7, 6525 ED Nijmegen, The Netherlands}
\author{W. Bras} \affiliation{Netherlands Organisation for Scientific Research (NWO), Dubble GRG, European
Synchrotron Radiation Facility, Grenoble, France}
\author{J.C. Maan}
\affiliation{High Field Magnet Laboratory, Institute for Molecules and Materials, Radboud University Nijmegen
Toernooiveld 7, 6525 ED Nijmegen, The Netherlands}
\author{P. C. M. Christianen}
\email{P.Christianen@science.ru.nl}
\affiliation{High Field Magnet Laboratory, Institute for Molecules and Materials, Radboud University Nijmegen
Toernooiveld 7, 6525 ED Nijmegen, The Netherlands}


\date{\today}

\begin{abstract}
We report the observation of a doubly-periodic surface defect-pattern in the liquid crystal 8CB, formed during the nematic--smectic~A phase transition. The pattern results from the antagonistic alignment of the 8CB molecules, which is homeotropic at the surface and planar in the bulk of the sample cell. Within the continuum Landau-deGennes theory of smectic liquid crystals, we find that the long period ($\approx$10 $\mu$m) of the pattern is given by the balance between the surface anchoring and the elastic energy of curvature wall defects. The short period ($\approx$1 $\mu$m) we attribute to a saddle-splay distortion, leading to a non-zero Gaussian curvature and causing the curvature walls to break up.
\end{abstract}


\maketitle


The richness of thermotropic liquid crystal (LC) phases~\cite{degennes:book}, their susceptibility to external fields and their unique optical properties make LCs  ideally suited to study symmetry breaking phase transitions. These transitions often involve the formation of isolated topological defects or complex ordered spatial structures (topological defect patterns)~\cite{chaikin:book}, with analogues in magnetism (grain boundaries, domain walls)~\cite{kleman:book}, superconductivity (Abrikosov lattice, stripes)~\cite{supercond} and cosmology (cosmic strings, monopoles)~\cite{lc:cosmology}. Contrary to cosmological or quantum systems, LC patterns can be  studied at room temperature using polarization microscopy, whence the formation, organization and kinetics of the defect structures can be fully explored.
Patterns in the LC nematic phase, with long range orientational order but no positional order, are mostly well understood and readily explained within a continuum elastic theory of LC~\cite{degennes:book}. In contrast, patterns in smectic LC phases are more difficult to describe due to the additional one-dimensional positional order. Many, sometimes rather complex, smectic patterns have been observed, such as undulations of the smectic layers in an applied magnetic field (Helfrich--Hurault instability)~\cite{degennes:book,kleman:book} or other periodic structures, like stripes~\cite{cladis:1975,lacaze:prl}, squares~\cite{lavrent:2006}, or hexagons~\cite{stewart:2003}. Usually, those structures are explained by the formation of focal conic domains or curvature walls, characterized by one typical length scale~\cite{kleman:book,blanc:1999,kleman:2000,ohzono:2012}. In this Letter we report the observation of a novel \emph{doubly-periodic} defect pattern, which is formed during the nematic-smectic A (N--SmA) phase transition of a liquid crystal in an applied magnetic field. The field imposes an orientation of the LC molecules in the bulk that is orthogonal to the preferred orientation at the surface of the sample cell. Most strikingly, the pattern has two distinct periods: a long one ($\approx$ 10 $\mu$m) along the field direction and a short one ($\approx$ 1 $\mu$m) perpendicular to the field. Interestingly, a quite similar texture develops in LC colloidal shells on cooling towards the  N--SmA phase transition \cite{per:2011}. We present a model describing the pattern using a geometric construction of a space-filling, energy minimizing, structure of equidistant (smectic) layers. Within this model we identify the driving mechanism as an elastic saddle-splay contribution ~\cite{barbero:2002,didonna:2002,me:2012} that breaks the symmetry in such a way that it naturally explains both distinct periodicities of the experiment and the orientation of the pattern with respect to the magnetic field direction.

\begin{figure}
\centering
\includegraphics[clip=true,width=\linewidth]{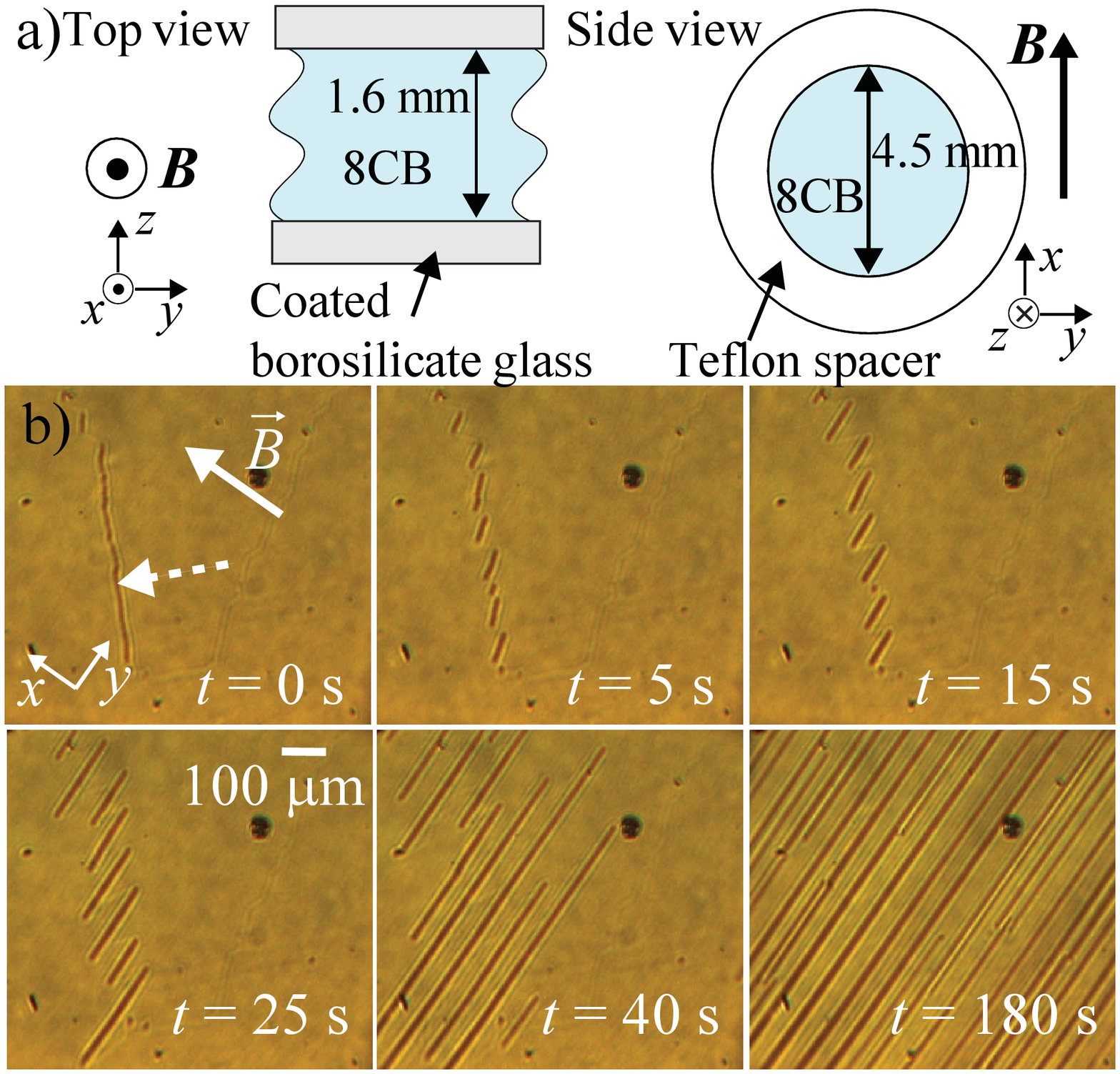}
\caption{\label{fig:cell} (Color) a) The sample cell consists of two borosilicate glass plates separated by a 4.5 mm diameter, 1.6 mm thick teflon
ring. The magnetic field $\vec{B}$ is applied in the plane of the cell (along the $x$-direction). b) Polarization microscopy images of the pattern formation during cooling (0.4 $^\circ$C/min) through the N--SmA phase transition of 8CB at 7 T.
At $t$ = 0 s ($T$ = 33.75~$^\circ$C) a line-defect is visible indicated by the dashed arrow. Upon further cooling this defect breaks up ($t$ = 5 s), aligns perpendicular to the field direction ($t$ = 15 s), and grows ($t$ = 25, 40 s). Finally the full sample surface is covered by stripes about 10--30~$\mu$m apart ($t$ = 180 s). }
\end{figure}

For our experiments we have used the liquid crystal 8CB (4-$n$-octyl-4'-cyanobiphenyl) which exhibits both a N and a SmA phase (SmA$\overset{33.5 ^\circ {\rm C}}{\longleftrightarrow}$ N $\overset{41.5 ^\circ {\rm C}}{\longleftrightarrow}$ I). The sample is contained in a cell consisting of two 0.4 mm
thick borosilicate glass plates, spaced by a teflon ring with 4.5~mm inner diameter and 1.6 mm
thickness (Fig. \ref{fig:cell}a). A 7~T static magnetic field $\vec{B}$ was applied in the plane of the cell, along the $x$-direction.
In-situ polarized microscopy was used to visualize the LC phase as a function of time.
The sample was positioned in the $xy$-plane in between two crossed polarizers at $\pm$ 45 $^\circ$ relative to $\vec{B}$. In this geometry the transmitted light intensity is maximal when the LC phase is aligned along $\vec{B}$ and minimal when the LC molecules are randomly aligned (isotropic phase), aligned along the viewing direction ($z$-axis) or aligned along one of the polarizer axes. 

Several cell glass coatings were used to vary the orientation of the 8CB molecules at the surface
and the strength of the surface anchoring. To obtain homeotropic alignment (parallel to the normal of the glass) with varying surface anchoring strength we used  hexamethyldisilazane (HMDS) coatings or
spin-coated polydimethylsiloxane (PDMS) layers, or the untreated glass. Alternatively, coating the cell with polyvinyl alcohol (PVA)
induces a planar molecular alignment. The following standard protocol was used: the sample was heated
to the isotropic phase for at least 10 minutes. Then after applying the magnetic field the sample was slowly cooled (0.4~$^{\circ}$C/min), through the magnetically aligned nematic phase, to a temperature within the smectic phase. When the LC pattern was fully developed the temperature was further decreased to room temperature and ${\bf B}$ was reduced to zero, after which the sample was taken out of the magnet to be investigated under a polarization microscope.

\begin{figure}
\centering
\includegraphics[clip=true,width=\linewidth]{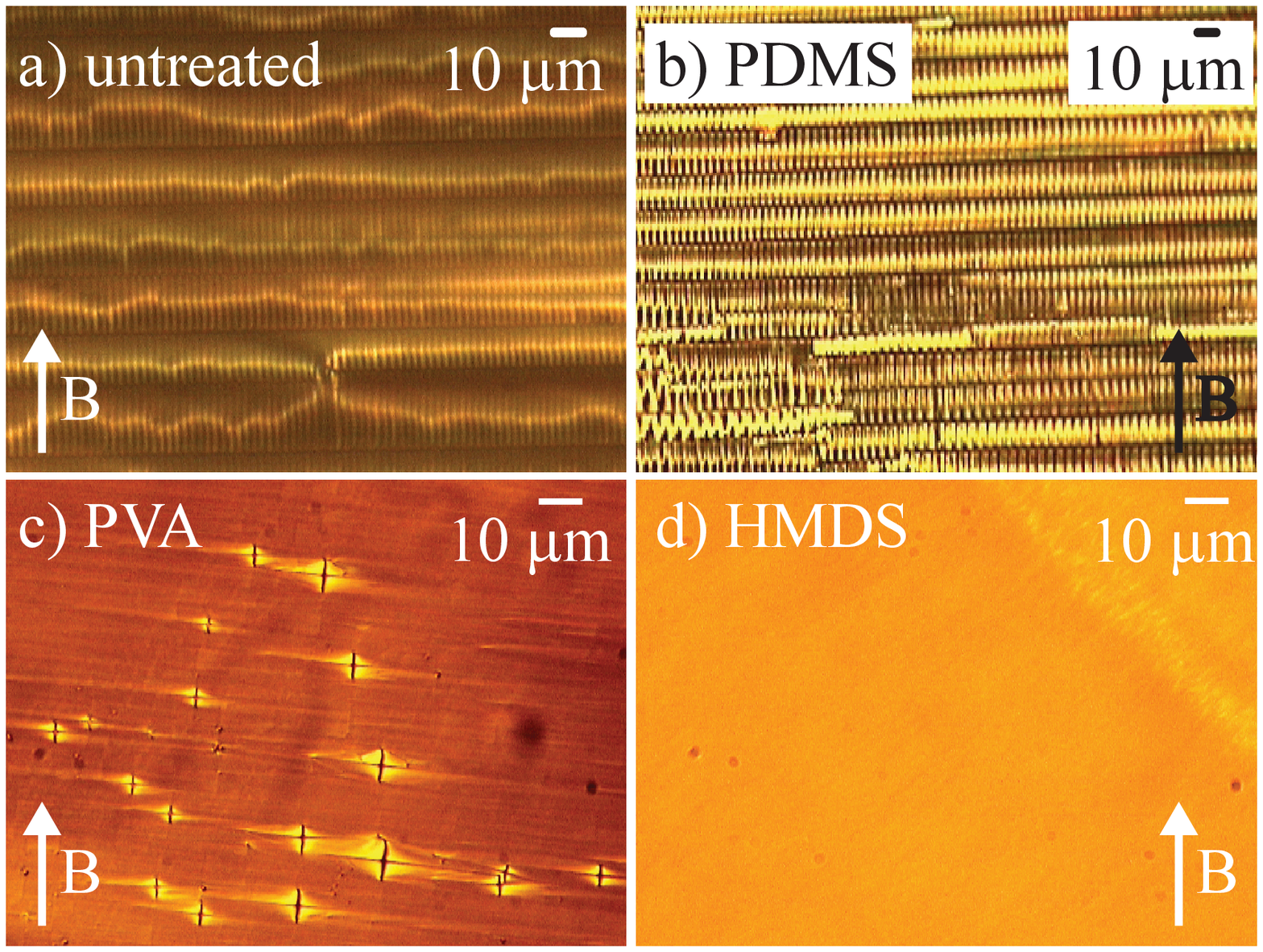}
\caption{\label{fig:images} (Color) A doubly-periodic surface pattern is visible for strong homeotropic surface anchoring, realized for a) untreated glass and b) PDMS coating. c) A PVA coating (planar surface alignment) leads to an aligned monodomain and focal conics. d) A HMDS coating gives an aligned monodomain due to the small homeotropic surface anchoring.}
\end{figure}

A typical experimental example of the pattern formation is shown in Fig.~\ref{fig:cell}b.
The first image ($t$ = 0 s, $T$ = 33.75~$^\circ$C) corresponds to a sample (untreated glass cell)
at 7~T close to the N--SmA phase transition. The overall transmitted light
intensity is high because in the bulk of the sample the LC phase is uniformly aligned. An elongated defect is visible as a black line (indicated by the arrow). Upon cooling, the defect breaks up in shorter lines ($t$ = 5 s), which rotate towards an orientation perpendicularly
to ${\bf B}$ ($t$ = 15 s). Subsequently the defects rapidly grow ($t$ = 25, 40 s).
Finally, the entire surface is covered by a surface pattern ($t$ = 180 s), consisting of many line defects oriented at 90$^\circ$
with respect to ${\bf B}$. The typical distance between the stripes is 10--30 $\mu$m. Heating the system through the SmA--N transition induces the reverse phenomenon: the line-defects melt and gradually disappear in
the nematic phase (not shown). Alternatively, cooling the surface pattern to room temperature
leads to a stable pattern that remains after the field is switched off and that can be studied under
the polarization microscope (Fig.~\ref{fig:images}a,b). The pattern is only present in the case of
homeotropic boundary conditions and relatively strong surface anchoring, realized for untreated glass
(Fig.~\ref{fig:images}a) and PDMS coating (Fig.~\ref{fig:images}b). When the surface anchoring is homeotropic
but too weak (HMDS, Fig.~\ref{fig:images}d) or for planar surface anchoring (PVA, Fig.~\ref{fig:images}c)
the pattern is absent.

Adjusting the focus of the microscope revealed that
the line patterns are formed on both the top and bottom surfaces of the cell, whereas the bulk
is homogeneously aligned. Most strikingly, the microscopy images at higher magnifications show
that the line pattern contains an additional fine structure: between the surfacelines smaller elongated
defects are visible with a periodicity of about 1--2~$\mu$m, perpendicular to the main pattern.
This secondary structure develops where the primary defect structure is
more dense and regular, i.e. in between straight defect lines. In contrast no secondary structure
is formed in the vicinity of an interrupted line pattern. This means that the secondary structure is most
clearly seen for the cell with strong surface anchoring (PDMS, (Fig. \ref{fig:images}b)), where a regular undulation
is observed in the inner structure of the line pattern, giving rise to a zig-zag pattern.

\begin{figure}
\centering
\includegraphics[clip=true,width=\linewidth]{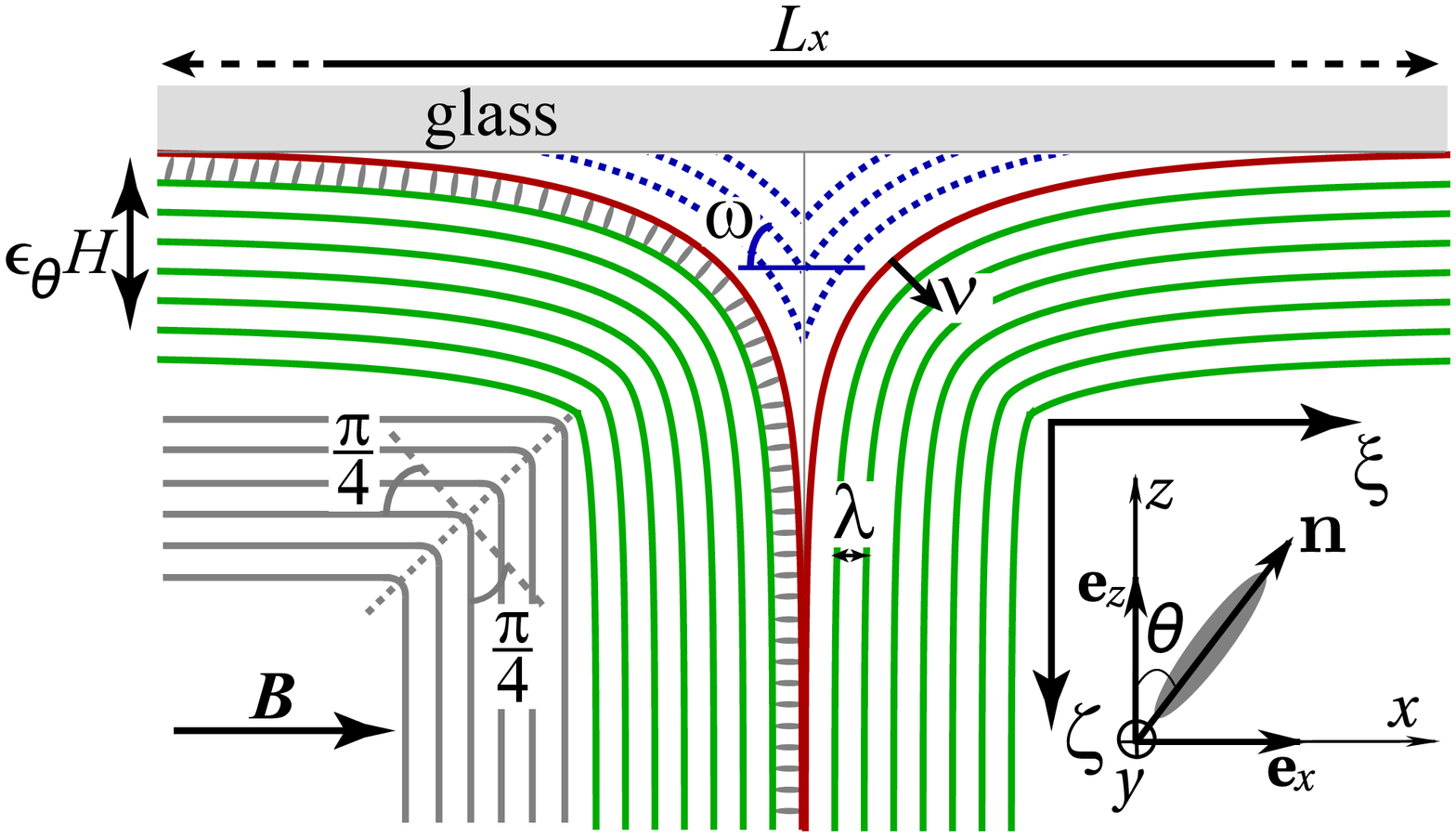}
\caption{\label{fig:layers} (Color) Part of the periodic structure close to the glass surface, consisting of  equidistant ($\lambda=\const$) smectic layers. The structure is defined by the (generating) red curve, given by $\zeta_0(\xi)$~\eqref{eq:z0} ($A\simeq1.1$). Space-filling is governed by a bending area (green green curves, \eqref{eq:layer}), a $\pi/2$-grain boundary (grey region) and a curvature wall at the surface (blue region). The resulting structure is periodic in the $x$-direction with period $L_x\gg \epsilon_\theta H$~\eqref{eq:lx}.
All further symbols are explained in the text.}
\end{figure}

We start our theoretical description in the nematic phase (above the transition temperature $T>T_c$ ($T_c\equiv T_{\rm N-SmA}$)), which is described by a unit vector $\bn$ called the director, pointing along the averaged orientation of the LC molecules. In the bulk the director is aligned parallel to the magnetic field ${\bf e}_x$, while close to the glass surface the molecules tend to align along the normal to the surface ${\bf e}_z$ (homeotropic anchoring, see Fig. \ref{fig:layers}). The director $\bn$ reorients by bending in order to minimize the sum of elastic and magnetic free energy:
\be\label{eq:fnem}
{\cal F}_{\rm nem}=\frac 12\int_V dV\, \{ K |\nabla \bn|^2 - \chi_aB^2 (\bn\cdot {\bf e}_x)^2\}.
\ee
We are far above the Freedericksz threshold $B_{\rm cr}=\frac \pi H\sqrt{\frac K{\chi_a}}\simeq 10^{-2}$~T~\cite{freed}, where  $K\simeq7\cdot 10^{-12}$~N is the elastic modulus, $\chi_a\simeq10^{-7}$ is the diamagnetic anisotropy in cgs units~\cite{degennes:book} and $2H=1.6$~mm is the thickness of the cell. The angle $\theta$ between the $z$-axis and the director $\bn=\sin\theta\,{\bf e}_x+\cos\theta\,{\bf e}_z$ varies along the thickness of the sample as~\cite{me:2012}:
\be\label{eq:theta}
\theta(\zeta)=\arcsin\bigg(\frac{A e^{2\zeta}-1}{A e^{2\zeta}+1}\bigg),\qquad A=\frac{1+\mu}{1-\mu}.
\ee
Here $\mu=\frac{\sqrt{K\chi_aB^2}}{W_a}$, $W_a$ is the anchoring strength associated with the energy cost $\frac 12 W_a\sin^2\theta|_{\zeta=0}$ (Rapini--Papoular form) for the deviation of $\bn$ from its preferred orientation along the normal to the surface. The dimensionless coordinate $\zeta =\frac{1-z/H}{\epsilon_\theta }$ measures the distance from the glass surface in the $z$-direction in units of the coherence length $\epsilon_\theta H =\sqrt{\frac K{\chi_aB^2}}\simeq 0.45~\mu$m.

The question we pose here is what happens when the SmA order, characterized by equally spaced layers, is imposed on this bent nematic structure. The smectic order is characterized by the complex order parameter $\rho e^{i\phi}$, where $\rho$ is the amplitude of the smectic density modulation, $\phi$ is the scalar function parametrizing  the smectic layers, so that $\nabla\phi$ is parallel to the layer's normal. For SmA the director coincides with the normal to the layers, such as  $\bn=\nabla\phi/|\nabla\phi|$. The associated Landau--deGennes free energy is~\cite{degennes:book}:
\be\label{eq:fsm}
{\cal F}_{\rm sm} = \!\int \!dV\bigg\{\frac C2 \big(\lambda^2|\nabla\rho|^2+\rho^2|\nabla\phi-\bn|^2\big) + \frac r2 \rho^2 +\frac g 4\rho^4\bigg\}.
\ee
For 8CB the compression modulus $C\simeq 10^6~$J/m$^3$ and the interlayer spacing $\lambda\simeq 3.2$~nm~\cite{lacaze:prl}. The coefficients $r$ and $g$ are regular Landau coefficients, where only $r$ depends on temperature, while $g>0$. We assume that the spatial variations of $\rho$ are negligible, because they occur within the characteristic lengthscale $\lambda\ll \epsilon_\theta H$ and prior to the change of orientation due to the cooling. Then $\rho=\sqrt{-\frac rg}\propto\sqrt{T_c-T}$ and the second  term in~\eqref{eq:fsm} is minimized if
\be\label{eq:equid}
\nabla\phi=\bn \quad\to\quad \frac{dx}{dz}=\cot\theta,
\ee
yielding the formation of equidistant smectic layers ($|\nabla\phi|=1$) on top of the distorted nematic phase. Substituting  $\theta$ from \eqref{eq:theta} and integrating \eqref{eq:equid} we find:
\be\label{eq:z0}
\zeta_0(\xi)=\frac 12\log\bigg(\frac{\coth (\xi/2)}A\bigg),
\ee
which parametrizes the smectic layer in terms of the dimensionless coordinate $\xi =x/(\epsilon_\theta H)$ along the $x$-axis as shown in Fig.~\ref{fig:layers} (red curve, $A\simeq1.1$). This important result for a closed form of the generating curve is  derived, based on the competition between anchoring and magnetic energy, mediated by elasticity~\eqref{eq:fnem}  under the global geometric constraint of equidistant layers~\eqref{eq:equid}, rather than assumed  {\it a priori}  as in refs.~\cite{parodi:1972,lacaze:prl,cladis:1975}. To obtain a space-filling two dimensional structure  we perform a parallel transport of the generating curve $\zeta_0(\xi)$ along its normal $\nu=\nu_x{\bf e}_x+\nu_z{\bf e}_z$ (see Fig.~\ref{fig:layers}), yielding a set of equidistant layers with the $j$th layer given by:
\be\label{eq:layer}
{\bf x}_j=(\xi+j\alpha \nu_x){\bf e }_x +(\zeta_0(\xi)+j\alpha \nu_z){\bf e}_z, \quad  j=0,\pm1,\pm2\ldots
\ee
Here $\alpha=\frac\lambda{\epsilon_{\theta}H}$ is the dimensionless interlayer distance,  the components of the normal $\nu_x= \frac{\csch \xi}{\sqrt{4+\csch ^2\xi}}$ and $\nu_z= \frac2{\sqrt{4+\csch ^2\xi}}$.  In the bulk ($\zeta\to \infty$) we obtain a set of flat parallel layers perpendicular to the $x$-axis.   The curved smectic layers (green lines in Fig.~\ref{fig:layers}) cannot fill the space without defects such as curvature walls shown by the grey and dashed-blue regions. Note that  Fig.~\ref{fig:layers}  illustrates only part of the whole structure, while the actual size of the grey region is determined by the energy  balance between the surface and the bulk contributions (see below).  Within our two dimensional model the size of the blue region is characterized  by the intersection of the two red curves (at the $z$-axis) at $\xi|_{\zeta=0}=\pm2 \coth^{-1}(A)$ and not by the energy minimizing structure of this region. This latter approach gives different space-filling structures as described previously~\cite{parodi:1972,lacaze:prl}.

The resulting structure consists of stripes parallel to the $y$-axis with period $L_x$ in the $x$-direction, along the magnetic field, like in the experiment. $L_x$ is determined by two contributions (see supplemental material for derivation~\cite{suppl}):
\be\label{eq:lx}
L_x\simeq 2\frac{W_a -2\rho\sqrt{KC}(1-\frac \pi4)}{\chi_a B^2}+2\frac {\epsilon_\theta H}{\max{|\kappa|}}.
\ee
The first term corresponds to the competition between the anchoring energy  (green region in~Fig. \ref{fig:layers}) and the energy of the curvature wall (grey region), given by  $f_\omega = \rho\sqrt{K C} (\tan \omega -\omega)\cos\omega$~\cite{blanc:1999} ($\omega=\pi/4$) per unit area. The second term originates from a non-zero curvature of the generating curve~\eqref{eq:z0}, given by $\kappa={\p_{\xi\xi}\zeta_0}/{(1+(\p_\xi\zeta_{0})^2)^{3/2}}$. The first term in~\eqref{eq:lx} is non-negative only if $W_a\gtrsim\rho\cdot10^{-3}$~J/m$^2$, yielding $L_x\simeq (9\pm4)~\mu$m for a strong surface anchoring ($W_a\simeq (3\pm1)\cdot 10^{-4}$~J/m$^2$) and a smectic order parameter $\rho\simeq 0.1$ in the vicinity of the N--SmA transition, which is compatible with our experimental observations. Within this model indeed no striped defect-patterns are expected for planar and small homeotropic surface anchoring.

The appearance of the second periodicity within the pattern is our most important experimental finding and to explain it is a truly challenging task. A plausible mechanism is the growth of the elastic constants in the vicinity of the N--SmA phase transition, in particular an enhanced  saddle-splay contribution~\cite{barbero:2002,didonna:2002}:
\be\label{eq:fK24}
{\cal F}_{K_{24}}=-K_{24}\int dV\,\big\{\nabla\cdot \big[\bn (\nabla\cdot \bn)+\bn\times\nabla\times\bn\big]\big\},
\ee
which should be added to~\eqref{eq:fnem}. Note that the expression under the integral can be cast into the form $\bn\cdot\nabla\times{\bf\Omega}$~\cite{kamien:rev}, where ${\bf\Omega}$ is the spin connection. This is  equivalent to the Gaussian curvature of the  smectic layers, since $\bn$ coincides with its normal or analogous to the Wen--Zee action~\cite{wenzee,wenzee1}, responsible for the curvature coupling in effective field theory of the quantum Hall systems. Thus layers with non-zero Gaussian curvature may become energetically favorable relative to the layers with zero Gaussian curvature considered above.

\begin{figure}[t]
\centering
\includegraphics[width=0.85\linewidth]{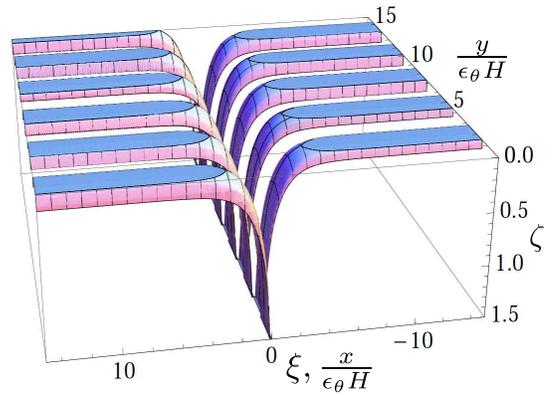}
\caption{\label{fig:2period} (Color online) The sum of the functions $\zeta_0(\xi)+\ve \hat\zeta(y/(\epsilon_\theta H))$ given by ~\eqref{eq:z0}, \eqref{eq:z1} and mirror reflection with respect to $x=0$, followed by a shift of $\pi$ along the $y$-axis. Calculated with $\ve=0.2$,  $\mu=0.06$, $\tau\simeq8$, $\omega\simeq2.4$ ($A\simeq1.1$).}
\end{figure}

To quantify the saddle-splay effect we perturb the nematic state~\eqref{eq:theta} by a small variation of polar $\hat\theta(y,z)$ and azimuthal $\hat\vp(y,z)$ angles. The linearised perturbation of the director $\bn$ then reads $\hat\bn=\hat\theta\cos\theta\,{\bf e}_x+\hat\vp\sin\theta\,{\bf e}_y-\hat\theta\sin\theta\,{\bf e}_z+O(\ve^2)$. Solving the variational problem for $\hat\theta=\ve f(z)\sin(q_y y)$ and $\hat\vp=\ve g(z)\cos(q_y y)$ ~\cite{me:2012}, we can find the threshold $(q_y,\mu)$ when the nematic state~\eqref{eq:theta} is linearly unstable with respect to periodic distortions along the $y$-direction. The incompressibility of the smectic layers~\eqref{eq:equid} implies that the linear correction to the level set function $\phi+\ve\hat\phi$ should satisfy  $\nabla\hat\phi=\nabla\hat\bn$, leading to:
\be\label{eq:z1}
\frac{dy}{dz}=-\frac{\hat\theta}{\hat\vp}\quad \to \quad
\sin (q_y y )=(\cosh \zeta+\mu\sinh \zeta)^{-\sigma}.
\ee
Here $\sigma=\frac{\omega(1+\mu\omega)}{\mu^2 \tau(\mu+\omega)}$, $\omega^2=1+q_y^2\epsilon_\theta ^2 H^2$ and $\tau=\frac{K_{24}}K$~\cite{me:2012}. The correction to  the level set function $\zeta_0(\xi)$~\eqref{eq:z0} at the onset of instability aquires a simple analytic form $\hat\zeta\simeq\cosh^{-1}\big[\sin (q_y y)^{-1/\sigma}\big]$ ($\mu\simeq0.06$  for strong anchoring  $W_a\simeq 3\cdot 10^{-4}$~J/m$^2$).  In Fig.~\ref{fig:2period} we plot $\zeta_0(\xi)+\ve\hat\zeta(y/(\epsilon_\theta H))$~\cite{note} for a periodically distorted generating layer (red curve in Fig.~\ref{fig:layers}) with non-zero Gaussian curvature.  This simplified analysis  allows to identify the intrinsic periodicity of modulated stripes $L_{y}\simeq1.5~\mu$m for $|K_{24}|\simeq 8K$, yielding a ratio $L_{x}/L_y\simeq 6$ compatible with our experimental observations. Indeed, `breaking' of the smectic layers in the vicinity of the glass surface  could relax the structure and prevent the formation of energetically expensive curvature wall defects  unavoidable in the planar case (blue region in Fig.~\ref{fig:layers}).

It is well known that curvature walls can become unstable with respect to focal conic domains (FCDs)~\cite{blanc:1999,kleman:2000}. FCDs are characterised by a semi-major axis $a$ of the ellipse and its eccentricity $e$. The latter can be related to the angle of a curvature wall as, $e=\sin(\omega/2)$~\cite{kleman:2000},  yielding $L_y/L_x=\sqrt{1-e^2}\simeq0.7$ ($\omega=\pi/2$), which is observed in experiments~\cite{ohzono:2012}. However, the formation of FCDs, with one characteristic length scale $L_x\simeq L_y$, is different from the appearance of our doubly-periodic pattern. 

In conclusion, we have observed a new \emph{doubly-periodic} defect texture in liquid crystals formed during the N--SmA phase transition. Based on a simple geometric argument and energy minimization including the saddle-splay elastic contribution, we have proposed a plausible scenario for instabilities at two different length scales. Our findings can  give insight into the  organization of other lamellar microstructures, ubiquitous in nature~\cite{cartwight:2007,vukusic:2003}, and can be of general interest to analyze symmetry breaking phase transitions in condensed matter systems~\cite{chaikin:book,supercond}.


\begin{acknowledgments}
O.V.M. acknowledges stimulating discussions with J.~Suorsa, G. Napoli, A. Fasolino and M. Katsnelson. G.T., W.B and P.C.M.C. acknowledge valuable discussions with Y.K. Levine and H.F.
Gleeson. This work was supported by EuroMagNET II under EU Contract No. 228043 and by the
Stichting voor Fundamenteel Onderzoek der Materie financially supported by the
Nederlandse Organisatie voor Wetenschappelijk Onderzoek.
\end{acknowledgments}

\bibliography{Manyuhina_ref}

\begin{thebibliography}{25}
\bibitem[{not()}]{note}
\bibinfo{note}{The amplitude of the perturbation~$\ve$ cannot be identified
  within a linear stability analysis.}
\end{thebibliography}

\end{document}